\documentclass[twocolumn]{aastex62} 
\usepackage{amsmath}
\usepackage{longtable, booktabs, threeparttablex}

\graphicspath{{./}{figures/}}

\shorttitle{Measuring Apsidal Clustering}
\shortauthors{Siraj, Chyba, \& Tremaine}

\begin{document}

\title{Measuring Apsidal Clustering}

\email{siraj@princeton.edu}

\author{Amir Siraj}
\affil{Department of Astrophysical Sciences, Princeton University, 4 Ivy Lane, Princeton, NJ 08544, USA}

\author{Christopher F. Chyba}
\affil{Department of Astrophysical Sciences, Princeton University, 4 Ivy Lane, Princeton, NJ 08544, USA}
\affil{School of Public and International Affairs, Princeton University, 20 Prospect Avenue, Princeton, NJ 08544, USA}

\author{Scott Tremaine}
\affil{School of Natural Sciences, Institute for Advanced Study, Princeton, NJ 08540, USA}

\begin{abstract}

The decade-long debate over the existence of apsidal clustering in the outer solar system is poised for reignition given the plethora of distant trans-Neptunian object (TNO) discoveries expected from the forthcoming Vera C. Rubin Observatory's Legacy Survey of Space and Time (LSST). Here, we present a new conditional-likelihood method to measure apsidal clustering that is insensitive to uneven survey footprints. We calculate the long-term orbital stability of distant TNOs, which allows us to expand the known sample of relevant objects from 21 to 25. We apply our new method to this up-to-date sample, showing that the significance of the apsidal clustering in the outer solar system has fallen from $2.7\sigma$ to $1.9\sigma$, and that the direction of clustering is not well constrained. This new method is suitable for application to the growing sample of known TNOs, and the results will reveal whether the evidence for a hypothetical Planet X from apsidal clustering is real or spurious.

\end{abstract}


\section{Introduction}

Are the eccentricity vectors of long-lived distant trans-Neptunian objects (TNOs) in the outer solar system distributed randomly, or are they clustered about a particular direction? This question has been at the center of the debate over a hypothetical `Planet X' or `Planet Nine' for the past decade \citep{2014Natur.507..471T, 2016AJ....151...22B, 2016AJ....152..221S, 2017AJ....154...50S, 2019AJ....157...62B, 2020PSJ.....1...28B,
2021AJ....162..219B,
2021PSJ.....2...59N,
2022ApJS..258...41B, 2025ApJ...978..139S}. The primary reason for the ongoing debate, aside from the uncertainties arising from small number statistics, is that it is difficult to distinguish whether perceived apsidal clustering genuinely reflects the intrinsic distribution of orbits, or if it is merely an artifact of complicated survey selection functions. With an imminent barrage of new distant TNO discoveries expected from the forthcoming  Legacy Survey of Space and Time (LSST) on the Vera C.\ Rubin Observatory \citep{2025AJ....170...99K}, a method that extracts the true apsidal clustering signal from a complicated combination of survey selection functions is needed now more than ever. Such a method can be constructed using the same principles introduced in the \cite{2025MNRAS.543L..27S} technique to measure the mean plane of the outer solar system.

In this work, we derive a new conditional-likelihood method for measuring apsidal clustering and demonstrate this method on the current census of distant, stable TNOs. In Section \ref{2D}, we introduce a simplified two-dimensional version of our method, and in Section \ref{3D} we describe the full, three-dimensional version. In Section \ref{mc}, we illustrate the efficacy of our method with Monte Carlo simulations. In Section \ref{app}, we provide an up-to-date sample of TNOs using the results of new stability simulations, and describe the differences between this sample and smaller samples used in previous works. In Section \ref{methodology}, we describe the apsidal clustering measurement methodologies we use. In Section \ref{results}, we report the results and disentangle the effects of our new measurement method from the effects
of our updated sample. Finally, in Section \ref{discussion}, we summarize our key findings and discuss implications of this work.

\section{2D Conditional-Likelihood Method}
\label{2D}

Here we present a 2D method to measure clustering in longitude of perihelion $(\varpi)$ that is insensitive to uneven survey footprints under the assumptions stated below. We use `2D' to mean that we assume all orbits have zero inclination and prograde motion, and are therefore coplanar in the ecliptic plane. We begin by describing the 2D method because it is more intuitive than the full 3D case, which allows the orbits to have nonzero inclinations.

We model the phase-space probability distribution of TNOs in canonical coordinates as
\begin{equation}
F(E,J,\varpi),
\end{equation}
where $E=\tfrac{1}{2}v^2-GM_\odot/r=-\tfrac{1}{2}GM_\odot/a$ is the energy, $J=|\mathbf r\times\mathbf v|=[GM_\odot a(1-e^2)]^{1/2}$ is the angular-momentum magnitude, $\varpi$ is the longitude of perihelion, $a$ is the semimajor axis, and $e$ is the eccentricity. If the distribution of $\varpi$ is independent of $E$ and $J$, we can write a separable form, 
\begin{equation}
\label{eq:sepF}
F(E,J,\varpi)\;=\;g(E,J)\,f(\varpi),
\end{equation}
where $g$ and $f$ are probabilities. The function $g$ is the orientation-free component (depending only on orbit size and shape through $E$ and $J$) and $f$ is the intrinsic directional distribution for $\varpi$. In the absence of apsidal clustering $f(\varpi)$ is a constant. Equation \eqref{eq:sepF} is an assumption about the population under study, not a general property of all TNO populations. This is the same kind of ensemble-level modeling assumption adopted in previous analyses of apsidal clustering. In particular, resonant populations can couple apsidal angle to orbital elements, so for such populations one would need a more general form of $F(E,J,\varpi)$ or a stratified analysis by dynamical class. Throughout this work we apply Equation \eqref{eq:sepF} only to the detached, distant population selected in Section \ref{app}, for which we seek a single ensemble-level distribution of $\varpi$ across the sample.

Any anisotropic pointing or seasonal effects -- or other selection effects that depend on where on the sky and at what distance an object was found -- are carried by the on-sky survey selection function $w(\mathbf r)$. The distance to the object is $r=|\mathbf r|$ and the ecliptic longitude (defining the direction of $\mathbf r$) is $\lambda$. Also,
\begin{equation}
\label{eq:lambda}
\lambda \;\equiv\; \varpi + \nu,
\end{equation}
where $\nu$ is the true anomaly. We therefore write the detected-state density for an object with $(\mathbf r,a,e,\varpi)$ as
\begin{equation}
w(\mathbf r)\,g(E,J)\,f(\varpi).
\end{equation}
We emphasize that writing the detected-state density in this form is not a statement that $r$, $a$, and $e$ are statistically independent; the correlations among these quantities are already encoded by orbital dynamics. The only assumption needed for the conditional likelihood is weaker: after conditioning on the observed $(\lambda,r,a,e)$, any residual selection factor common to the two admissible values of $\varpi$ cancels. The distance $r$ is related to the true anomaly $\nu$ as follows,
\begin{equation}
r=\frac{a(1-e^2)}{1+e\cos \nu}.
\end{equation}
At fixed $(r,a,e)$ this sets $|\nu|=\phi$ where
\begin{equation}
\label{eq:phi} 
\phi=\cos^{-1}\frac{a(1-e^2)/r-1}{e},
\end{equation}
with $0\le\phi\le\pi$. Thus the two allowed states are $\nu=\pm\phi$. By Eq. \eqref{eq:lambda}, the allowed longitudes of perihelion at fixed $(\lambda,r,a,e)$ are then,
\begin{equation}
\label{eq:allowed}
\varpi \in \{\lambda-\phi,\;\lambda+\phi\}.
\end{equation}
In steady state this equal weighting is the Keplerian expectation because
\begin{equation}
\frac{dt}{d\nu}\propto \frac{1}{(1+e\cos \nu)^2}
\end{equation}
which is an even function of $\nu$. The two solutions $\nu=\pm\phi$ therefore have equal intrinsic phase-space weight at fixed $(r,a,e)$. Our additional assumption is only that the residual detection efficiency at fixed geometry does not depend on the sign of the radial velocity. For object $i$, conditioning on the measured $(\lambda_i,r_i,a_i,e_i)$, and not on $\varpi_i$, the per-object conditional probability, $p\,(\varpi \mid \lambda_i,r_i,a_i,e_i)$, is
\begin{align}
\label{eq:cond}
\frac{w(\mathbf r_i)\,g(E_i,J_i)\,f(\varpi)}
{w(\mathbf r_i)\,g(E_i,J_i)\,f(\lambda_i-\phi_i)+w(\mathbf r_i)\,g(E_i,J_i)\,f(\lambda_i+\phi_i)},
\end{align}
where $\phi_i$ is computed from $(r_i,a_i,e_i)$ using Eq.~\eqref{eq:phi}. Since $w(\mathbf r_i)$ and $g(E_i,J_i)$ do not depend on $\varpi$ at fixed $(\lambda_i,r_i,a_i,e_i)$, they cancel, yielding
\begin{equation}
\label{eq:condsimple_pt1}
p(\varpi \mid \lambda_i,r_i,a_i,e_i)
=\frac{f(\varpi)}{Z_i},\qquad
\end{equation}

\begin{equation}
\label{eq:condsimple_pt2}
Z_i \equiv f(\lambda_i-\phi_i)+f(\lambda_i+\phi_i).
\end{equation}
Thus the footprint $w$ and any orientation-free factor common to the two admissible solutions drop out of the conditional directional likelihood.

We adopt the von Mises form for the intrinsic distribution of $\varpi$ because it is standard and algebraically convenient for modeling a distribution about a circle:
\begin{equation}
\label{eq:vm}
f(\varpi)=\exp(\kappa\cos(\varpi-\mu)),
\end{equation}
with preferred direction $\mu$ and concentration $\kappa$. Here $\kappa=0$ corresponds to a uniform distribution, and larger $\kappa$ implies stronger clustering. In the small-angle limit, $\exp[\kappa\cos(\Delta\varpi)]\sim \exp(\kappa)\exp[-\kappa(\Delta\varpi)^2/2]$, so the characteristic angular width is of order $\kappa^{-1/2}$ radians. Substituting Eq.~\eqref{eq:vm} into Eq.~\eqref{eq:condsimple_pt1} yields the per-object normalizer,
\begin{align}
\label{eq:Ci}
Z_i
&=\exp(\kappa\cos(\lambda_i-\mu-\phi_i))
+\exp(\kappa\cos(\lambda_i-\mu+\phi_i))\nonumber\\
&=2\,\exp(\kappa\cos(\lambda_i-\mu)\cos\phi_i)\nonumber \\&\quad\quad \times
\cosh(\kappa\sin(\lambda_i-\mu)\sin\phi_i).
\end{align}

For a sample $\{(\varpi_i,\lambda_i,r_i,a_i,e_i)\}_{i=1}^N$, the likelihood and log-likelihood are
\begin{equation}
\label{eq:logL_1}
L(\mu,\kappa)=\prod_{i=1}^N \frac{1}{Z_i}\,\exp(\kappa\cos(\varpi_i-\mu)),
\end{equation}
\begin{equation}
\label{eq:logL_2}
\log L(\mu,\kappa)=\sum_{i=1}^N\Bigl[\kappa\cos(\varpi_i-\mu)-\log Z_i\Bigr].
\end{equation}
We maximize $\log L$ over $(\mu,\kappa)$ to estimate the preferred perihelion direction and concentration. A na\"ive method (one that does not attempt to distinguish between observational selection effects and the intrinsic distribution of orbits) adopting the same functional form as in Equation \eqref{eq:vm} is described in Appendix \ref{sec:2D_naive}, allowing for a direct comparison in Section \ref{mc}.

One might ask whether this method is sensitive to survey selection effects that depend on the direction or magnitude of the proper motion. Selection on proper motion can, in principle, distinguish the two solutions $\nu=\pm\phi$, but we can show that the effect is negligible for distant TNOs. Let $\mathbf V_\perp$ be the sky-plane component of the relative velocity $\mathbf v-\mathbf V_\oplus$ (object minus Earth). Flipping $\nu$ changes only the sign of the radial speed $v_r$, which alters the sky-plane speed by $|\Delta \mathbf V_\perp|=2|v_r|\sin\alpha \lesssim 2|v_r|(1 \, \mathrm{AU} / r)$, where $\alpha$ is the small Sun--object--Earth angle. Thus the fractional change in sky-plane speed is
\begin{equation}
\frac{\Delta V_\perp}{V_\perp}=\frac{|\Delta \mathbf V_\perp|}{|\mathbf V_\perp|}\;\lesssim\;\frac{2|v_r|}{|\mathbf V_\perp|}\,\frac{\mathrm{1 \,AU}}{r}.
\end{equation}
For typical distant TNOs, $|\mathbf V_\perp|\sim 30~\mathrm{km\,s^{-1}}$ (Earth's reflex motion), $|v_r|\lesssim 1~\mathrm{km\,s^{-1}}$, and $r\gtrsim 40~\mathrm{AU}$, giving $\Delta V_\perp/V_\perp\lesssim 2\times 10^{-3}$ (i.e. $0.2\%$), which is negligible. Any corresponding change in apparent brightness over the timescale relevant for discovery and follow-up is likewise expected to be negligible for such distant objects.

\section{3D Conditional-Likelihood Method}
\label{3D}

We extend the method in Section \ref{2D} from the planar case ($\varpi$) to the full 3D perihelion direction given by the unit eccentricity (Runge--Lenz) vector $\hat{\mathbf e}$. The steady-state population model becomes
\begin{equation}
\label{eq:sepF3d}
F(E,J,\hat{\mathbf e},\hat{\mathbf J})\;=\;g(E,J,\hat{\mathbf J})\,f(\hat{\mathbf e}\mid \hat{\mathbf J}),
\end{equation}
where $g$ depends on $(E,J,\hat{\mathbf J})$ and $f(\hat{\mathbf e}\mid \hat{\mathbf J})$ is the intrinsic directional distribution within the orbital plane. As in Section \ref{2D}, any on-sky selection is carried by $w(\mathbf r)$, and we assume no residual selection on $\hat{\mathbf e}$ once the geometry is fixed. This factorization is an assumption about the detached, distant population analyzed here rather than a universal description of all TNO populations. If the perihelion-direction distribution depends explicitly on $(E,J)$ within the population of interest, the model should be generalized accordingly. The unit orbital pole is
\begin{equation}
\hat{\mathbf J}\;\equiv\;\frac{\mathbf r\times\mathbf v}{|\mathbf r\times\mathbf v|}.
\end{equation}
For an object observed at $(\hat{\mathbf r},r,a,e)$, Eq.~\eqref{eq:phi} implies two possible true anomalies $\nu=\pm\phi$. In 3D the allowed perihelion directions are the two rotations of $\hat{\mathbf r}$ within the orbital plane by angle $\phi$ about $\hat{\mathbf J}$,
\begin{equation}
\label{eq:ehatpm3d}
\hat{\mathbf e}_\pm\;=\;\hat{\mathbf r}\cos\phi\;\pm\;(\hat{\mathbf J}\times\hat{\mathbf r})\sin\phi,
\end{equation}
where $\hat{\mathbf J}\!\cdot\!\hat{\mathbf r}=0$. We again take these two possibilities to have equal intrinsic prior weight in steady state, since the corresponding $\nu=\pm\phi$ solutions have equal Keplerian phase-space weight. As in 2D, the only additional assumption is that the residual detection efficiency at fixed geometry does not depend on the sign of the radial velocity.

Conditioning on the measured $(\hat{\mathbf r}_i,r_i,a_i,e_i,\hat{\mathbf J}_i)$ for object $i$, and not on $\hat{\mathbf e}_i$, the per-object conditional directional likelihood is
\begin{equation}
\label{eq:condsimple3d_pt1}
p\,(\hat{\mathbf e}\mid \hat{\mathbf r}_i,r_i,a_i,e_i,\hat{\mathbf J}_i)
=\frac{f(\hat{\mathbf e}\mid \hat{\mathbf J}_i)}{Z_i},
\end{equation}
\begin{equation}
\label{eq:condsimple3d_pt2}
Z_i \;\equiv\; f(\hat{\mathbf e}_{i,+}\mid \hat{\mathbf J}_i)\;+\;f(\hat{\mathbf e}_{i,-}\mid \hat{\mathbf J}_i),
\end{equation}
where the footprint $w$ and any orientation-free factor common to the two admissible solutions have canceled out as in 2D (for fixed $i$, $g$ does not vary across the two admissible $\hat{\mathbf e}_{i,\pm}$).

We adopt a von Mises--Fisher weight within the orbital plane for the intrinsic distribution of $\hat{\mathbf e}$,
\begin{equation}
\label{eq:vmf3d}
f(\hat{\mathbf e}\mid \hat{\mathbf J}) =\exp(\kappa\,\hat{\mathbf k}\!\cdot\!\hat{\mathbf e}) \, \delta(\hat{\mathbf e}\!\cdot\!\hat{\mathbf J}),
\end{equation}
with preferred direction $\hat{\mathbf k}$ and concentration $\kappa\ge 0$. As in 2D, $\kappa=0$ corresponds to an isotropic in-plane distribution and larger $\kappa$ implies stronger concentration about $\hat{\mathbf k}$. The kinematic constraint that $\hat{\mathbf e}$ is perpendicular to $\hat{\mathbf J}$ is enforced by the delta function $\delta(\hat{\mathbf e}\!\cdot\!\hat{\mathbf J})$.\footnote{The delta function
$\delta(\hat{\mathbf e}\cdot\hat{\mathbf J})$
in Eq. \eqref{eq:vmf3d} simply encodes that only in-plane directions with $\hat{\mathbf e}\cdot\hat{\mathbf J}=0$ contribute. Formally, putting $u=\hat{\mathbf e}\cdot\hat{\mathbf J}$ one
has 
\[
\int_{-1}^{1}\delta(u)\,du \;=\; 1.
\] 
After conditioning on $(\hat{\mathbf r}_i,r_i,a_i,e_i,\hat{\mathbf J}_i)$, only the two admissible directions $\hat{\mathbf e}_{i,\pm}$ remain, so the delta functions cancel between the numerator and \textbf{$Z_i$} in Eqs.~\eqref{eq:condsimple3d_pt1}--\eqref{eq:condsimple3d_pt2}.
} Substituting Eq.~\eqref{eq:vmf3d} into Eq.~\eqref{eq:condsimple3d_pt1} and using Eq.~\eqref{eq:ehatpm3d} gives the per-object normalizer
\begin{align}
\label{eq:Ci3d}
Z_i
&=\exp(\kappa\,\hat{\mathbf k}\!\cdot\!\hat{\mathbf e}_{i,+})
+\exp(\kappa\,\hat{\mathbf k}\!\cdot\!\hat{\mathbf e}_{i,-})\nonumber\\
&=2\,\exp\bigl(\kappa\,(\hat{\mathbf k}\!\cdot\!\hat{\mathbf r}_i)\,\cos\phi_i\bigr)\,
\cosh\bigl(\kappa\,(\hat{\mathbf k}\!\cdot\!(\hat{\mathbf J}_i\times\hat{\mathbf r}_i))\,\sin\phi_i\bigr),
\end{align}
where $\phi_i$ is computed from $(r_i,a_i,e_i)$ via Eq.~\eqref{eq:phi}.

For a sample $\{(\hat{\mathbf e}_i,\hat{\mathbf r}_i,r_i,a_i,e_i,\hat{\mathbf J}_i)\}_{i=1}^N$, the likelihood and log-likelihood are
\begin{equation}
\label{eq:logL3d_1}
L(\hat{\mathbf k},\kappa)=\prod_{i=1}^N \frac{1}{Z_i}\,\exp(\kappa\,\hat{\mathbf k}\!\cdot\!\hat{\mathbf e}_i),
\end{equation}
\begin{equation}
\label{eq:logL3d_2}
\log L(\hat{\mathbf k},\kappa)=\sum_{i=1}^N\Bigl[\kappa\,\hat{\mathbf k}\!\cdot\!\hat{\mathbf e}_i-\log Z_i\Bigr].
\end{equation}
We maximize $\log L$ over $(\hat{\mathbf k},\kappa)$ to estimate the preferred 3D perihelion direction and concentration. A na\"ive method adopting the same functional form as in Equation \eqref{eq:vmf3d} is described in Appendix \ref{sec:3D_naive}, allowing for a direct comparison in Section \ref{mc}.

\section{Monte Carlo Validation}
\label{mc}

To test whether our conditional-likelihood methods work as expected, we run two Monte Carlo experiments: a 2D experiment that models clustering in perihelion longitude $\varpi$, and a 3D experiment that models clustering in the perihelion-direction unit vector $\hat {\mathbf e}$ on the sphere. In each trial we draw a large intrinsic population, ``observe'' it with a two–field survey subject to a heliocentric distance limit, and down–select to a fixed number of detections per survey. The conditional-likelihood estimators themselves are described in Sections \ref{2D} and \ref{3D}, and the na\"ive estimators are described in Appendices \ref{sec:2D_naive} and \ref{sec:3D_naive}; here we detail the population and survey generation.

In both the 2D and 3D cases, the synthetic population consists of orbits with semimajor axes drawn randomly from a uniform distribution in the range $200 - 500 \mathrm{\; AU}$, perihelia from a uniform distribution in the range $30 - 80 \mathrm{\; AU}$ and mean anomaly from a uniform distribution in the range $0 - 2\pi$.

\begin{figure}
 \centering
\includegraphics[width=\linewidth]{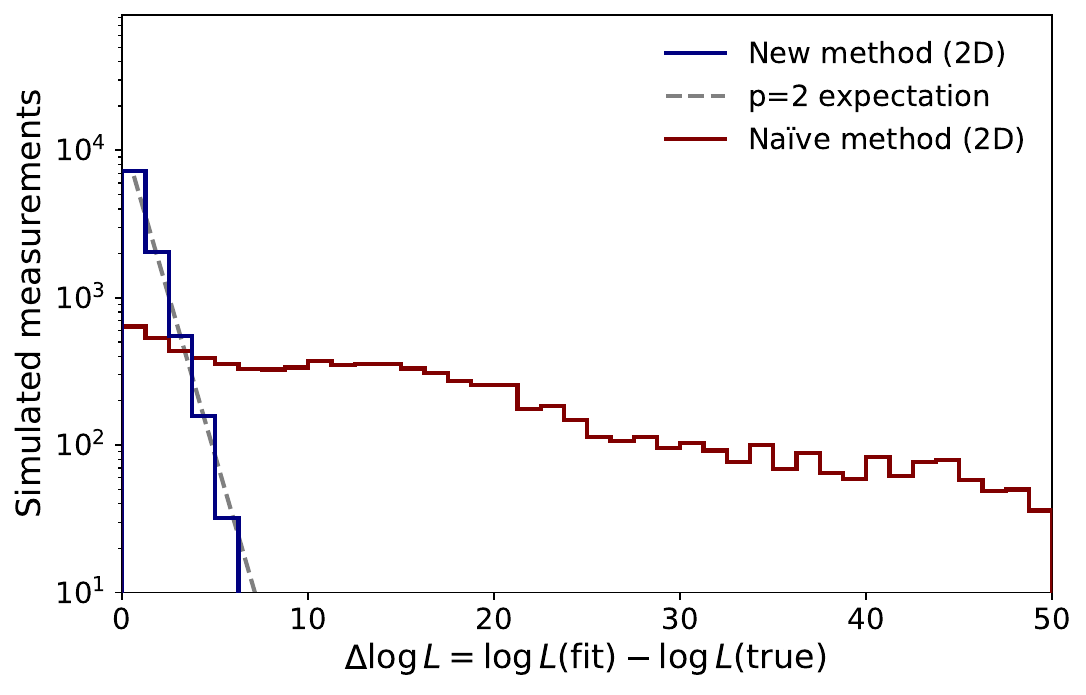}
\caption{Distribution of $\Delta \log{L}$ for the Monte Carlo simulation described in Section \ref{mc}, in the 2D case.}
\label{fig:2D_mc}
\end{figure}

In the 2D case, the longitude of perihelion $\varpi$ is then chosen from a von Mises distribution (see Eq. \eqref{eq:vm}) with some preferred direction $\mu$ and concentration parameter $\kappa$. The results displayed in Figure \ref{fig:2D_mc} correspond to a test with $\kappa = 1$. Each hypothetical survey is composed of two ecliptic–longitude pointings of width $20^{\circ}$; the two field-center longitudes are drawn independently from a uniform distribution in the range $0 - 2\pi$. An object is ``detectable'' if it lies within the union of these longitude windows and satisfies the observability cutoff $r < 100 \mathrm{\; AU}$. We randomly select 100 detectable candidates per survey. We repeat this procedure over $10^4$ independently initialized surveys to build up ensembles of simulated measurements.

\begin{figure}
 \centering
\includegraphics[width=\linewidth]{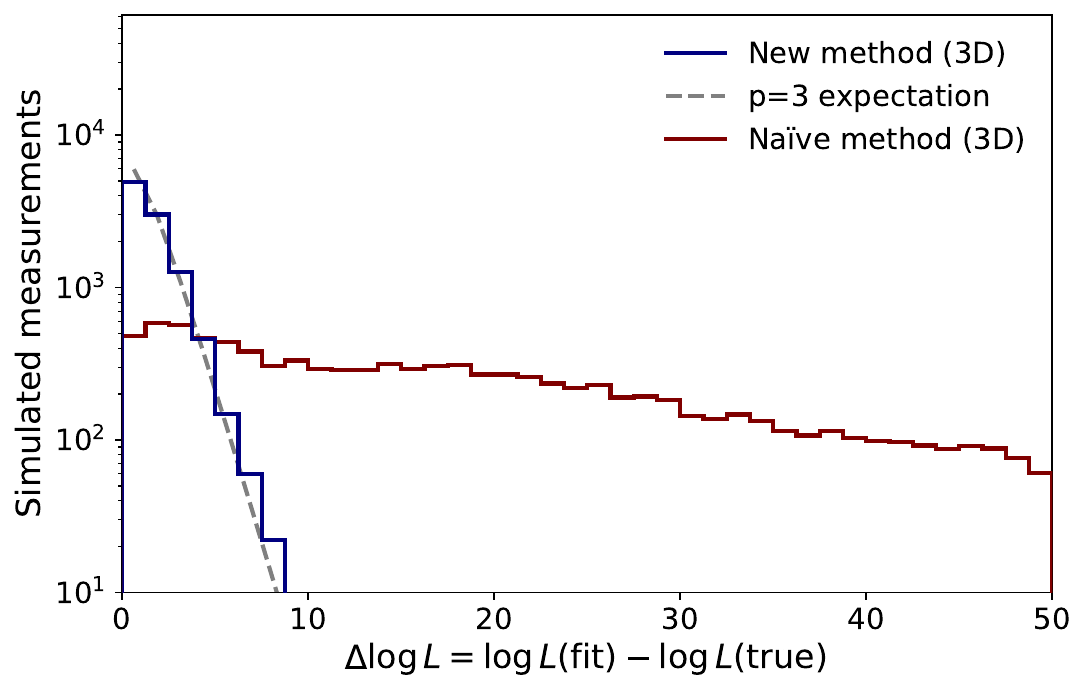}
\caption{Distribution of $\Delta \log{L}$ for the Monte Carlo simulation described in Section \ref{mc}, in the 3D case.}
\label{fig:3D_mc}
\end{figure}

In the 3D case, after drawing the semimajor axis, perihelion, and mean anomaly, we draw the perihelion–direction unit vector $\hat{\mathbf e}$ from a von Mises–Fisher distribution on the sphere (see Eq. \eqref{eq:vmf3d}) with some preferred direction $\hat{\mathbf k}$ and concentration parameter $\kappa$. Conditional on $\hat{\mathbf e}$, we choose an orbital pole $\hat{\mathbf J}$ uniformly on the circle perpendicular to $\hat{\mathbf e}$; thus, we do not assume any explicit inclination distribution relative to the ecliptic. We are then able to calculate the instantaneous sky-direction unit vector $\hat{\mathbf r}$ and record the ecliptic longitude and latitude $(\lambda, \beta)$. The results displayed in Figure \ref{fig:3D_mc} correspond to a test with $\kappa = 1$. Each hypothetical survey is composed of two $20^{\circ} \times 20^{\circ}$ pointings for which the central ecliptic longitude ($\lambda$) of each pointing is drawn from a uniform distribution in the range $0 - 360^{\circ}$, and the central ecliptic latitude ($\beta$) of each pointing is drawn from a uniform distribution in $\sin(\beta)$ for $\mid \beta \mid \leq 20^{\circ}$. An object is ``detectable'' if its $(\lambda,\beta)$ falls inside either field and $r < 100\mbox{ AU}$. As in the 2D case, we randomly select 100 detectable objects per survey, and we repeat this procedure over $10^4$ independently initialized surveys.

The resulting detections are then passed to the estimators: in the 2D case, Appendix \ref{sec:2D_naive} for the na\"ive estimator and Section \ref{2D} for the conditional-likelihood estimator, in the 3D case, these are described in Appendix \ref{sec:3D_naive} and Section \ref{3D}, respectively. Finally, for the fits produced by the estimator for each set of detections, we calculate the $\Delta \log{L}$ between the log-likelihood of best-fit parameters and the log-likelihood of the parameters used to build the true population. The results are displayed in Figures \ref{fig:2D_mc} and \ref{fig:3D_mc}, illustrating that the new method performs vastly better than the na\"ive estimators. Under a correctly specified likelihood, $2\Delta \log L$ should be a $\chi^2$ distribution with $p$ parameters, so the $\Delta \log L$ histograms for the new method should have probability density equal to $[1/\Gamma(p/2)] (\Delta \log L)^{(p/2 - 1)} e^{-\Delta \log L}$. We verified that the results from the new method match these expectations ($p = 2$ for the 2D case, $p = 3$ for the 3D case). The gray lines overlaid on Figures \ref{fig:2D_mc} and \ref{fig:3D_mc} illustrate the $p=2$ and $p=3$ expectations, respectively.

\section{Sample}
\label{app}

Since the writing of \cite{2025ApJ...978..139S}, 23 additional objects with barycentric semimajor axes in the range $90 - 1000\mbox{\;AU}$ and pericenter $q > 30\mbox{\;AU}$, with $U_{MPC} < 7$ have been added to the census.\footnote{See \url{http://www.minorplanetcenter.org/iau/info/UValue.html}. The MPC ``uncertainty parameter'' $U_{MPC}$ is denoted the ``condition code'' in the JPL database.}  We followed the stability simulation procedure outlined in \cite{2025ApJ...978..139S} to determine the classification of each new object as stable, metastable, or unstable, using the same definitions as in \cite{2021AJ....162..219B} and \cite{2025ApJ...978..139S}. Thus, for each object we create a set of $200$ clones, by drawing from the full orbital covariance matrix about the nominal solution as reported by the JPL Small-Body Database\footnote{\url{https://ssd.jpl.nasa.gov/tools/sbdb_query.html}}. We then integrate the orbit of each clone for $4$ Gyr, and determine the stability of the object based on the fraction of surviving clones and the degree to which the semimajor axes of the surviving clones have changed. We carry out all $n$-body simulations in this work using the \texttt{REBOUND} package \citep{2012A&A...537A.128R} with the \texttt{MERCURIUS} hybrid symplectic integrator \citep{2019MNRAS.485.5490R}. 

The simulations revealed that 9 of the 23 additional TNOs were stable or metastable. This brings the total number of known stable or metastable TNOs with semimajor axes in the range $90 - 1000\mbox{\;AU}$ to 60. We applied the new conditional-likelihood apsidal clustering measurement methods (described in Sections \ref{2D} and \ref{3D}) while enforcing lower limits on semimajor axis in increments of $10 \mathrm{\;AU}$ in an effort to locate the onset of clustering, which is expected at large semimajor axes in the presence of a Planet X. We calculated the significance of the clustering as $Z=\sqrt{2\,\Delta\log L}$, where $\Delta\log L=\log L(\hat{k}_{\mathrm{ML}},\hat{\kappa}_{\mathrm{ML}})-\log L(\kappa=0)$. For both the set of 60 stable and metastable objects known at the time of this work and the 51 metastable objects known at the time of \cite{2025ApJ...978..139S}, we find that a lower bound of $a_{\rm min} = 160\mbox{\;AU}$ produces the most significant result. This is consistent with the conclusion of \cite{2025ApJ...978..139S}, who reported the first measurement of the semimajor axis of clustering onset at $\sim 170\mbox{ AU}$ with negligible statistical difference between $a_{\rm min} = 170\mbox{ AU}$ and $a_{\rm min} = 160\mbox{ AU}$. Note that the entire \cite{2021AJ....162..219B} sample has semimajor axes that exceed this threshold.

As such, the most up-to-date sample (which we shall refer to as the SCT26 sample) comprises the 25 stable and metastable objects with $a > 160 \mbox{ AU}$. The SCT25 sample is the 21-object subset of this group that appears in \cite{2025ApJ...978..139S}. Finally, the \cite{2021AJ....162..219B}, or BB21, sample is a further subset of 11 objects. The four objects in the SCT26 sample that are not present in the SCT25 sample are 2017 OF201 \citep{2025arXiv250515806C}, 2021 RR205 \citep{2022MPEC....M...118M}, 2023 KQ14 \citep{2025NatAs...9.1309C}, and 2024 FX26 \citep{2025MPEC....M...20S}.\footnote{2017 OF201, 2021 RR205, and 2023 KQ14 (but not 2024 FX26) would have been included in the BB21 sample if they had been discovered by the time of the writing of \cite{2021AJ....162..219B}.} The distribution of $\varpi$ for the three samples is shown in Figure \ref{fig:samples}.

\section{Methodology}
\label{methodology}

The conditional-likelihood procedures described in Sections \ref{2D} and \ref{3D} return estimates of the clustering direction and the degree of concentration about that direction. We can also quantify the uncertainty in the preferred direction of clustering. We do so by profiling the conditional likelihood over the angle of interest while maximizing over all remaining parameters. In 2D the parameter of interest is the preferred longitude of perihelion $\mu$. For all possible values of $\mu$, we maximize the 2D conditional log–likelihood over the concentration parameter $\kappa\ge 0$, and record the maximized log-likelihood as a function of $\mu$. In 3D the parameter of interest is the ecliptic longitude of the preferred direction, $k_{\rm lon}$ ($\varpi$); analogously, at each fixed $k_{\rm lon}$ we maximize over latitude $k_{\rm lat}$ and $\kappa\ge 0$ and record the resulting log-likelihood as a function of $k_{\rm lon}$. We then find the values of $\mu$ and $k_{\rm lon}$ that correspond to the 1-sigma uncertainty envelope by identifying where the difference between the global maximum $\log{L}$ and the maximum $\log{L}$ given those values of $\mu$ and $k_{\rm lon}$ is $0.5$.

In addition, we wish to compare the results from the 2D and 3D conditional-likelihood methodologies to a simple method that estimates the clustering direction and strength of the clustering based only on the $\varpi$ values and using a Monte Carlo procedure, in order to illustrate how the results change when the conditional-likelihood methods are used. This is the same estimator as the 2D na\"ive estimator presented in Appendix \ref{sec:2D_naive}, the only difference here is presentational: here we assess significance by Monte Carlo under the uniform null, much like in \cite{2019AJ....157...62B} and \cite{2021AJ....162..219B}, whereas in Appendix \ref{sec:2D_naive} the same information is captured in the von Mises fit.

For this simple method, we compute
\begin{equation}
\bar{s}=\frac{1}{N}\sum_{i=1}^N \sin\varpi_i,\qquad
\bar{c}=\frac{1}{N}\sum_{i=1}^N \cos\varpi_i,
\end{equation}
and define the clustering angle and magnitude as 
\begin{equation}
\hat{\mu}=\mathrm{atan2}(\bar{s},\bar{c}),\qquad
R=\sqrt{\bar{s}^2+\bar{c}^2}.
\end{equation}
Here $\hat{\mu}$ is the average direction of the unit vectors $(\cos\varpi_i,\sin\varpi_i)$ and $R$ is the length of their mean resultant vector.

We also quantify the angular dispersion about the average direction using the circular standard deviation,
\begin{equation}
\sigma_{\mathrm{circ}} \;=\; \sqrt{-2\ln R}.
\end{equation}

To assess significance under the null hypothesis of uniformly distributed angles, we perform a Monte Carlo test that preserves the sample cardinality. For each random trial, we draw a number of random angles equal to the number of actual objects in the sample from a uniform distribution between 0 and $2\pi$ and compute the length of the mean resultant vector, $R_{MC}$. The $p$-value is then simply the fraction of random trials for which $R_{MC} > R$. We then convert the Monte Carlo $p$–value to a one-sided Gaussian significance by inverting the standard normal cumulative distribution function, allowing for an easy comparison to the significances reported from the conditional-likelihood methods.

\section{Results}
\label{results}

\begin{figure}
 \centering
\includegraphics[width=\linewidth]{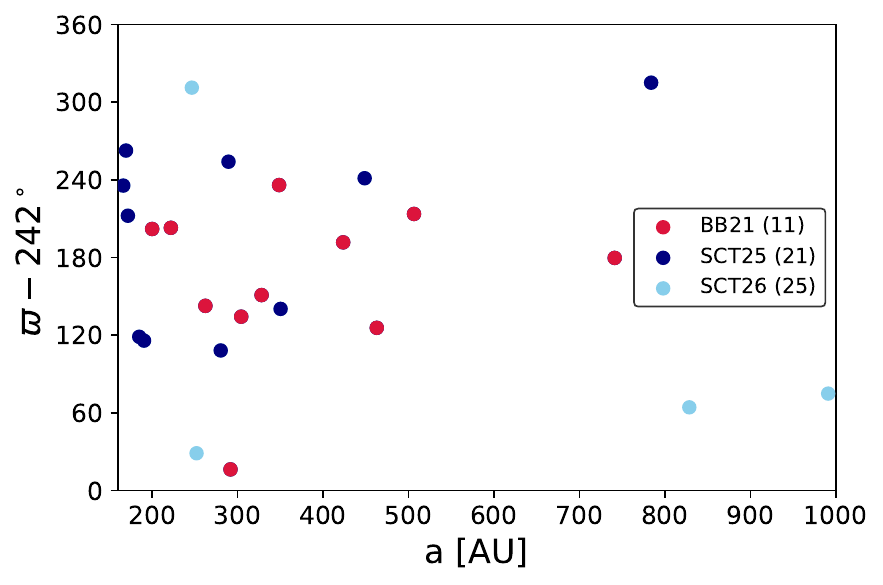}
\caption{Distribution of $\varpi$ for all three samples: BB21, SCT25, and this work (SCT26). BB21 is a subset of SCT25 and SCT25 is a subset of SCT26.}
\label{fig:samples}
\end{figure}

The results presented in Table \ref{tab:results} show the clustering significance and direction for the simple, 2D conditional-likelihood, and 3D conditional-likelihood methods, applied to the BB21, SCT25, and SCT26 samples, allowing us to disentangle the effects of the updated TNO sample from those of the new conditional-likelihood technique for measuring apsidal clustering. These results show that the SCT25 sample strengthened the evidence for apsidal clustering relative to the BB21 sample, but the latest additions in the SCT26 weakened it to below the level of the BB21 sample. The evidence for apsidal clustering now sits near $\sim 2\sigma$. In addition, they illustrate that while the direction of clustering according to the simple method (the circular mean) does not change significantly across the three samples, the circular standard deviation as well as the marginalized results from the conditional-likelihood method indicate that the direction of clustering is actually highly uncertain ($151^{+34}_{-59}\mbox{ deg}$ in longitude of perihelion).

\begin{table}
	\centering
	\caption{Apsidal clustering measurement results. As described in Section \ref{app}, the BB21 sample \citep{2021AJ....162..219B} has $N = 11$ TNOs, while the SCT25 sample \citep{2025ApJ...978..139S} has $N = 21$ and the SCT26 sample (this work) has $N = 25$. Both the significance of the clustering relative to the null, and the direction of the clustering in $\varpi$ are listed for each method applied to each sample.}
	\label{tab:results}
	\begin{tabular}{lccr}
		\hline
		Method  & BB21 & SCT25 & SCT26\\
		\hline
            Simple & $2.5\sigma$ & $2.8\sigma$ & $1.6\sigma$\\
             & $58\pm 52\mbox{ deg}$ & $67\pm 66\mbox{ deg}$ & $58\pm 84\mbox{ deg}$\\
            2D C.-L. & $2.2\sigma$  & $2.7\sigma$ & $1.7\sigma$\\
             & $41^{+52}_{-19}\mbox{ deg}$ &  $126^{+28}_{-35}\mbox{ deg}$ & $143^{+34}_{-55}\mbox{ deg}$\\
            3D C.-L. & $2.4\sigma$  & $2.7\sigma$ & $1.9\sigma$\\
             & $43^{+33}_{-16}\mbox{ deg}$ &  $125^{+30}_{-37}\mbox{ deg}$ & $151^{+34}_{-59}\mbox{ deg}$\\
		\hline
	\end{tabular}
\end{table}

Of the four objects included in the SCT26 sample that were not present in the SCT25 sample, the addition of two of them, namely 2021 RR205 and 2024 FX26, leaves both the 2D and 3D conditional-likelihood significances nearly unchanged (only by $0.1\sigma$ for both). The addition of the other two objects, 2017 OF201 and 2023 KQ14 (the two light blue points with coordinates $(64^{\circ},\, 829\mbox{\,AU})$ and $(29^{\circ}, \,252\mbox{\,AU})$ in Figure \ref{fig:samples}), results in the marked drop in statistical significance between SCT25 and SCT26.

The decrease in significance can be decomposed into, 1. the direct contributions of the two added objects evaluated at the SCT26 maximum likelihood estimate, and 2. a re-optimization effect on the other 23 objects (the 23-object subset of SCT26 evaluated at the SCT26 maximum likelihood estimate rather than its own). We find that in the 3D conditional-likelihood case, roughly half of the drop in the log-likelihood comes directly from the two objects (roughly a quarter from each) and the other half from the worsened fit to the original 23 when the global parameters are re-fit. No significant change between SCT25 and SCT26 was due to any of the best estimates for the orbital elements in the SCT25 data changing due to further observations.

\section{Discussion}
\label{discussion}

In this work, we introduced a new method for measuring apsidal clustering whose principal advantage is that uneven on-sky footprint factors cancel analytically under the population-model assumptions in Equations \eqref{eq:sepF} and \eqref{eq:sepF3d}, together with the solution-symmetry assumptions stated in Sections \ref{2D} and \ref{3D}. We validated the method with Monte Carlo simulations, updated the sample of relevant TNOs by applying stability simulations to the current census of objects, and applied our new method to the updated sample. With our full 3D method, we found that the evidence for apsidal clustering relative to the null (no clustering) has dropped from $2.7\sigma$ to $1.9\sigma$ significance due to additional stable and metastable objects admitted to the sample since \cite{2025ApJ...978..139S}. In addition, we found that the direction of clustering is only constrained to about half of the sky by the current sample of 25 objects. Thus the on-sky probability density distributions that have been used to search for a Planet X \citep{2021AJ....162..219B, 2024AJ....167..146B, 2025ApJ...978..139S} are too narrow. The present method should be viewed as complementary to, rather than a replacement for, detailed survey simulators, such as Sorcha \citep{2025AJ....170..100M, 2025AJ....170...97H}. The method is intended for pooled samples from multiple surveys, provided that any residual selection at fixed geometry is symmetric between the two admissible solutions. Within the stated model, however, the results in Table \ref{tab:results} imply that our new method does not sacrifice statistical power while removing uneven-footprint effects analytically. If a substantial solution-dependent asymmetry at fixed geometry were demonstrated in a characterized survey, then the resulting likelihood-ratio significance should be interpreted within that stated model and complemented by detailed survey simulations. This method therefore allows a footprint-independent test, under the assumptions in Equations \eqref{eq:sepF} and \eqref{eq:sepF3d} together with the solution-symmetry assumptions stated in Sections \ref{2D} and \ref{3D}, of whether apsidal clustering exists in the outer solar system as it is applied to the growing sample of TNOs in the forthcoming LSST era. One result should be to shed light on the evidence for the hypothetical Planet X.

\section*{Acknowledgements}

We are pleased to acknowledge that the work reported in this paper was substantially performed using the Princeton Research Computing resources at Princeton University which is consortium of groups led by the Princeton Institute for Computational Science and Engineering (PICSciE) and Office of Information Technology's Research Computing. We thank the anonymous referee for insightful comments that greatly improved the quality of the manuscript.


\appendix
\section{2D Na\"ive Method}
\label{sec:2D_naive}

To allow for a comparison in the Monte Carlo validation (see Section \ref{mc}) against the conditional-likelihood 2D method in Section \ref{2D}, we describe a na\"ive fit that uses only the observed longitudes of perihelion $\{\varpi_i\}_{i=1}^N$ and ignores the survey selection. We model
\begin{equation}
\label{eq:vm1d_kernel}
f(\varpi)=\exp(\kappa\cos(\varpi-\mu)),
\end{equation}
Normalizing $f$ on $[0,2\pi)$ gives
\begin{equation}
p(\varpi\mid\mu,\kappa)
=\frac{f(\varpi)}{\displaystyle\int_{0}^{2\pi}\! f(\theta)\,d\theta}
=\frac{\exp(\kappa\cos(\varpi-\mu))}
       {\displaystyle\int_{0}^{2\pi}\!\exp(\kappa\cos\theta)\,d\theta}
=\frac{\exp(\kappa\cos(\varpi-\mu))}{2\pi\,I_0(\kappa)}\;,
\end{equation}
where $I_0$ is a modified Bessel function. For $N$ independent measurements $\{\varpi_i\}_{i=1}^N$, the likelihood and log-likelihood are
\begin{equation}
L_{\mathrm{naive}}(\mu,\kappa)
=\prod_{i=1}^N \frac{\exp(\kappa\cos(\varpi_i-\mu))}{2\pi I_0(\kappa)}
=\frac{\exp(\kappa\sum_{i=1}^N \cos(\varpi_i-\mu))}{(2\pi I_0(\kappa))^N},
\end{equation}
\begin{equation}
\label{eq:logL1d_short}
\log L_{\mathrm{naive}}(\mu,\kappa)
=\sum_{i=1}^N \kappa\cos(\varpi_i-\mu)\;-\;N\log I_0(\kappa)\;+\;\text{const}.
\end{equation}
We define the vector sum and its mean length
\begin{equation}
\label{eq:resultant_short}
C\equiv\sum_{i=1}^N \cos\varpi_i,\qquad
S\equiv\sum_{i=1}^N \sin\varpi_i,\qquad
R\equiv\sqrt{C^2+S^2},\qquad
\bar R\equiv \frac{R}{N}.
\end{equation}
Maximizing the log-likelihood gives 
\begin{align}
\label{eq:muhat_naive_short}
\mu_{\mathrm{naive}}&=\operatorname{atan2}(S,\,C),\\
\label{eq:kappahat_naive_short}
\bar R&=\;\;\frac{I_1(\kappa_{\rm naive})}{I_0(\kappa_{\rm naive})}.
\end{align}

\section{3D Na\"ive Method}
\label{sec:3D_naive}

To allow for a comparison in the Monte Carlo validation (see Section \ref{mc}) against the conditional-likelihood 3D method in Section \ref{3D}, we give the na\"ive fit that uses only the observed unit eccentricity directions $\{\hat{\mathbf e}_i\}_{i=1}^N$ and ignores the survey selection. We model
\begin{equation}
\label{eq:vmf3d_naive_kernel}
f(\hat{\mathbf e})=\exp(\kappa\,\hat{\mathbf k}\!\cdot\!\hat{\mathbf e}),
\end{equation}
and normalize $f$ over the unit sphere,
\begin{equation}
\label{eq:vmf3d_norm}
p(\hat{\mathbf e}\mid\hat{\mathbf k},\kappa)
=\frac{f(\hat{\mathbf e})}{\displaystyle\int\! f(\hat{\mathbf u})\,d\Omega(\hat{\mathbf u})}
=\frac{\exp(\kappa\,\hat{\mathbf k}\!\cdot\!\hat{\mathbf e})}
       {\displaystyle\int_{0}^{2\pi}\!\!\int_{0}^{\pi}\exp(\kappa\cos\theta)\sin\theta\,d\theta\,d\phi}
=\frac{\kappa\,\exp(\kappa\,\hat{\mathbf k}\!\cdot\!\hat{\mathbf e})}{4\pi\,\sinh\kappa}\;.
\end{equation}
For $N$ independent measurements $\{\hat{\mathbf e}_i\}_{i=1}^N$, the likelihood and log-likelihood are
\begin{equation}
\label{eq:logL3d_naive}
L_{\mathrm{naive}}(\hat{\mathbf k},\kappa)
=\prod_{i=1}^N \frac{\kappa\,\exp(\kappa\,\hat{\mathbf k}\!\cdot\!\hat{\mathbf e}_i)}{4\pi\,\sinh\kappa}
=\frac{\kappa^{\,N}}{(4\pi\,\sinh\kappa)^{N}}\,
\exp(\kappa\,\hat{\mathbf k}\!\cdot\!\sum_{i=1}^N\hat{\mathbf e}_i),
\end{equation}
\begin{equation}
\log L_{\mathrm{naive}}(\hat{\mathbf k},\kappa)
=\kappa\,\hat{\mathbf k}\!\cdot\!\sum_{i=1}^N\hat{\mathbf e}_i
\;+\;N\log\kappa\;-\;N\log\sinh\kappa\;+\;\text{const}.
\end{equation}
We define the vector sum and its mean length
\begin{equation}
\label{eq:resultant3d_naive}
\mathbf R\equiv\sum_{i=1}^N \hat{\mathbf e}_i,\qquad
R\equiv|\mathbf R|,\qquad
\bar R\equiv \frac{R}{N}.
\end{equation}
Maximizing the log-likelihood gives,
\begin{align}
\label{eq:muhat3d_naive}
\hat{\mathbf k}_{\mathrm{naive}}&=\frac{\mathbf R}{|\mathbf R|},\\
\label{eq:kappahat3d_naive}
\bar R&=\coth\kappa_{\rm naive}\;-\;\frac{1}{\kappa_{\rm naive}}.
\end{align}

\newpage

\bibliography{bib}{}
\bibliographystyle{aasjournal}

\end{document}